\documentclass{PoS}
\usepackage{amsmath}
\usepackage{pb-diagram}
\usepackage{braket}

\title{Deconfining temperatures in SO(N) and SU(N) gauge theories}

\ShortTitle{Deconfining temperatures in SO(N) and SU(N) gauge theories}

\author{\speaker{Richard Lau}\footnote{Funded by the Science and Technology Facilities Council.} \ and Michael Teper\\
        Rudolf Peierls Centre for Theoretical Physics, University of Oxford\\
        E-mail: \email{richard.lau@physics.ox.ac.uk} \\
 		E-mail: \email{m.teper1@physics.ox.ac.uk}}

\abstract{We present our current results for the deconfining temperatures in $SO(N)$ gauge theories in 2+1 dimensions. $SO(2N)$ theories may help us to understand QCD at finite chemical potential since there is a large-$N$ orbifold equivalence between $SO(2N)$ QCD-like theories and $SU(N)$ QCD, and $SO(2N)$ theories do not have the sign problem present in QCD. We show that the deconfining temperatures in these two theories match at the large-$N$ limit. We also present results for ${SO(2N+1)}$ gauge theories and compare results for $SO(6)$ with $SU(4)$ gauge theories, which have the same Lie algebras but different centres.}

\FullConference{The 32nd International Symposium on Lattice Field Theory,\\
		23-28 June, 2014\\
		Columbia University New York, NY}

\begin{document}

\section{Introduction}

$SO(N)$ gauge theories do not have a fermion sign problem \cite{cherman11}, are orbifold equivalent to $SU(N)$ QCD \cite{unsal06}, and share a common large-$N$ limit with $SU(N)$ gauge theories \cite{lovelace82}. 
Some $SO(N)$ gauge groups also share Lie algebra equivalences with $SU(N)$ gauge groups such as $SO(4)\sim SU(2)\times SU(2)$ or $SO(6)\sim SU(4)$. 
All this indicates that we could investigate $SU(N)$ QCD at finite chemical potential through considering the equivalent $SO(N)$ gauge theories. 

There is a large-$N$ orbifold equivalence between $SO(2N)$ QCD-like theories and $SU(N)$ QCD \cite{cherman11}. 
This equivalence holds if we take the large-$N$ limit while relating the couplings $g$ in the two theories by $\left. g^{2} \right|_{SU(N \rightarrow \infty)} =  \left. g^{2} \right|_{SO(2N \rightarrow \infty)}$. 
Using this result, along with knowing that the leading correction between finite $N$ and the large-$N$ limit is $\mathcal{O}(1/N)$ for $SO(2N)$ and $\mathcal{O}(1/N^2)$ for $SU(N)$, we can construct a possible path connecting $SO(N)$ and $SU(N)$ gauge theories. 
\begin{align}
\begin{diagram}
\dgARROWLENGTH=6em
\node{SU(N \rightarrow\infty)} \arrow{e,t,<>}{\text{large-$N$ equivalence}} \arrow{s,l,<>}{\mathcal{O}\left(\frac{1}{N^2}\right)\text{ corrections}} 
    \node{SO(2N \rightarrow\infty)} \arrow{s,r,<>}{\mathcal{O}\left(\frac{1}{N}\right)\text{ corrections}} \\
\node{SU(N)} 			\node{SO(2N)} 
\end{diagram}
\label{eqn:equiv}
\end{align}

We showed at Lattice 2013 that we obtain the same large-$N$ limits for the string tension and mass spectrum from $SO(2N)$ and $SU(N)$ gauge theories in $D=2+1$ \cite{lau13}. 
In this contribution, we calculate $SO(N)$ deconfining temperatures in $2+1$ dimensions and we will show that they match $SU(N)$ values between Lie algebra equivalences and at the large-$N$ limit.

As before, we consider $D=2+1$ values because the $SO(N)$ $D=3+1$ bulk transition occurs at very small lattice spacings so that the volumes needed are currently too large to simulate \cite{forcrand03}.
However, in $D = 2 + 1$, the bulk transition occurs at larger lattice spacings and we can obtain continuum extrapolations at reasonable volumes \cite{bursa13}.
We use the standard plaquette action for an $SO(N)$ gauge theory.
\begin{align}
S=\beta\sum_{p}\left( 1-\frac{1}{N} \text{tr} (U_p) \right) \qquad\qquad \beta=\frac{2N}{ag^2} 
\end{align}

\section{Deconfinement}

We expect $SO(N)$ gauge theories to deconfine at some temperature $T=T_c$, just like $SU(N)$ gauge theories.
We can look for the deconfinement temperature by using an `order parameter' $O$ such as the `temporal' Polyakov loop $ \overline{l_P}$ \cite{liddle}.
The expectation value of the Polyakov loop $\braket{\overline{l_p}}$ is not invariant under a transformation with a non-trivial element of the centre. 
Hence, for gauge theories with non-trivial centres, such as $\mathbb{Z}_2$ $SO(2N)$, the expectation value is zero.
This corresponds to confinement, while a non-zero expectation value corresponds to deconfinement.
Hence, deconfinement corresponds here to a spontaneous breakdown of the $\mathbb{Z}_2$ symmetry.

\begin{figure}
	\centering
  	\includegraphics[width=0.8\textwidth]{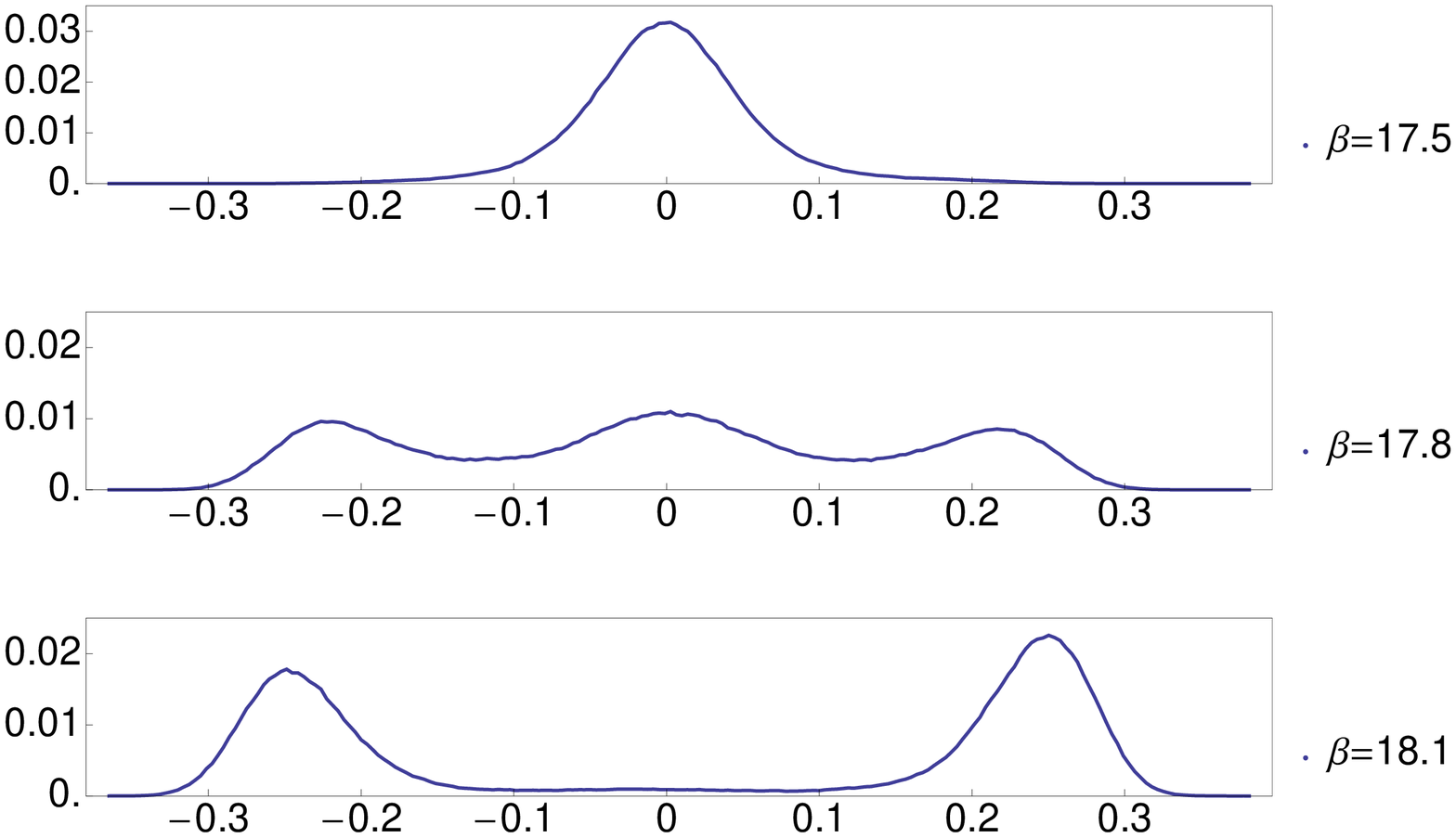} 
   	\caption{Histograms of $\braket{\overline{l_P}}$ in an $SO(6)$ $20^23$ volume at several values of $\beta$ around $\beta_c$.
	}
	\label{fig:so6-histogram}
\end{figure}

Using this order parameter, we can look for signs of the deconfining phase transition such as changes in the histogram peaks of the order parameter over a full configuration run.
We display an example of such histograms in Figure~\ref{fig:so6-histogram} for an $SO(6)$ $20^23$ volume.
As we increase $\beta$, we can see that a primary peak around zero disappears while two secondary peaks at non-zero values appear.
The primary peak represents the confined phase while the secondary peaks represent the deconfined phase.
The transition between the two states indicates that the $\beta$ range we considered is around $\beta_c$.
Furthermore, the coexistence of the two phases that we can see in the middle histogram indicates that the $SO(6)$ deconfining phase transition is first order.

\begin{figure}
	\centering
  	\includegraphics[width=0.8\textwidth]{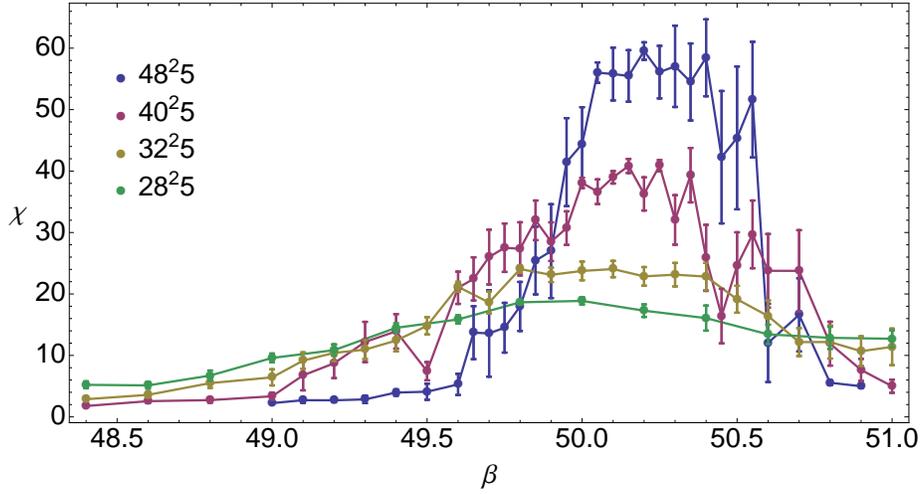} 
   	\caption{Susceptibility plot for $SO(8)$ $L_t=5$ volumes.}
	\label{fig:so8-susceptibility}
\end{figure}

\begin{figure}
	\centering
  	\includegraphics[width=0.8\textwidth]{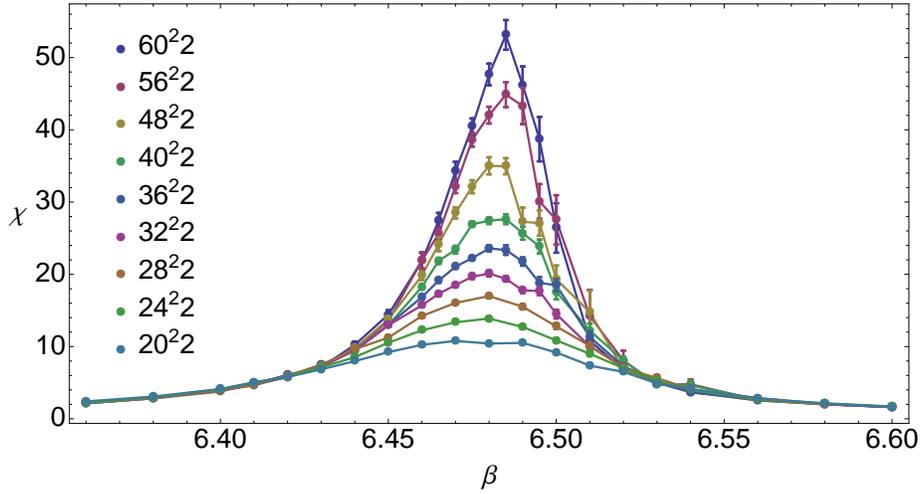} 
   	\caption{Susceptibility plot for $SO(4)$ $L_t=2$ volumes.}
	\label{fig:so4-susceptibility}
\end{figure}

We can construct a `susceptibility' $\chi_{\left| \overline{l_P}\right|}$ for the Polyakov loop.
\begin{align}
\chi_{\left| \overline{l_P}\right|}\sim \Braket{\left| \overline{l_P}\right|^2}-\Braket{\left| \overline{l_P}\right|}^2
\end{align}
We can calculate the susceptibility for different $\beta$ in the region of $\beta_c$.
The peak in this susceptibility when plotted against $\beta$ corresponds to $\beta_c$.
The peak structure can also indicate the order of the phase transition when we vary the finite spatial volume $V$.
As $V$ increases, the peak height increases as the peak converges towards the non-analyticity associated with the continuum phase transition, but the characteristic width changes depending on the transition's order.
For first order transitions, the characteristic width decreases at the same rate as the peak height increases so that the peak converges to a delta function.
We can see this in Figure~\ref{fig:so8-susceptibility} for an $SO(8)$ phase transition.
For second order transitions, the characteristic width decreases at a slower rate than the peak height increases so that the peak converges to a divergence.
We can see this in Figure~\ref{fig:so4-susceptibility} for an $SO(4)$ phase transition.

\begin{figure}
	\centering
  	\includegraphics[width=0.8\textwidth]{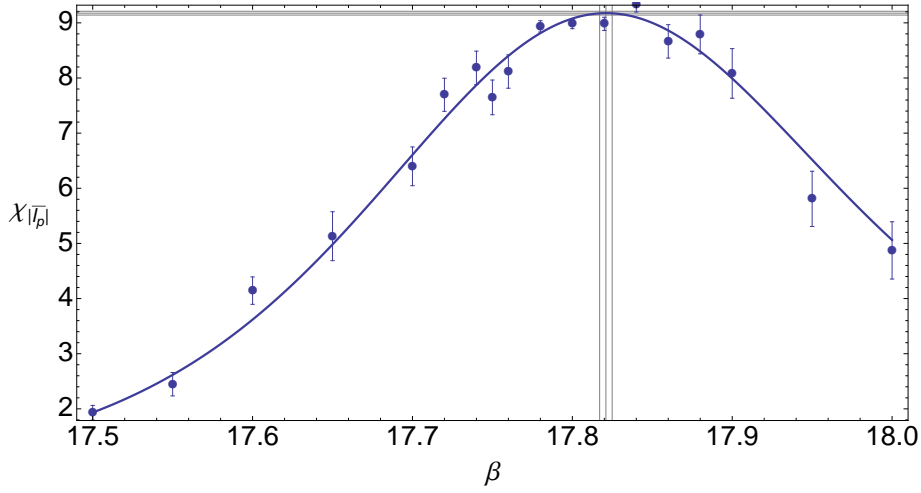} 
   	\caption{The susceptibility for an $SO(6)$ $20^23$ volume with reweighted results.}
	\label{fig:so6-peak}
\end{figure}

To identify accurately $\beta_c$ from the susceptibility peak, we use reweighting \cite{ferrenberg89}.
The principle behind reweighting is that we can consider the generation of lattice configurations as sampling an underlying density of states, which is independent of $\beta$.
If we could reconstruct the density of states, then we could calculate observables at an arbitrary value of $\beta$.
Reweighting allows us to calculate $\beta_c$ very accurately as we can see in Figure~\ref{fig:so6-peak}.

\section{$SO(N)$ measurements}

\begin{figure}
	\centering
  	\includegraphics[width=0.8\textwidth]{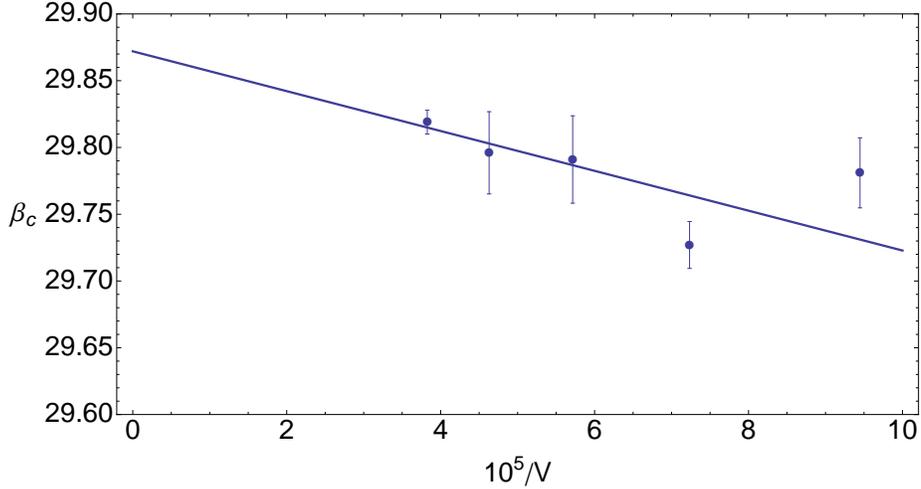} 
   	\caption{Infinite volume extrapolation for $SO(6)$ $L_t=6$ volumes.}
	\label{fig:so6-infinitevolume}
\end{figure}

For a fixed `temporal' length $L_t$, we calculate $\beta_c(V)$ for different spatial volumes $V$.
By using known results from finite size scaling, we can extrapolate $\beta_c(V)$ values for different spatial volumes $V$ to the infinite spatial volume limit $V\rightarrow\infty$.
For first order transitions, $\beta_c(V)$ varies linearly with $1/V$, as we can see in Figure~\ref{fig:so6-infinitevolume}.
For second order transitions, $\beta_c(V)$ varies with $1/V$ in a way determined by the critical exponents of the phase transition.

\begin{figure}
	\centering
  	\includegraphics[width=0.8\textwidth]{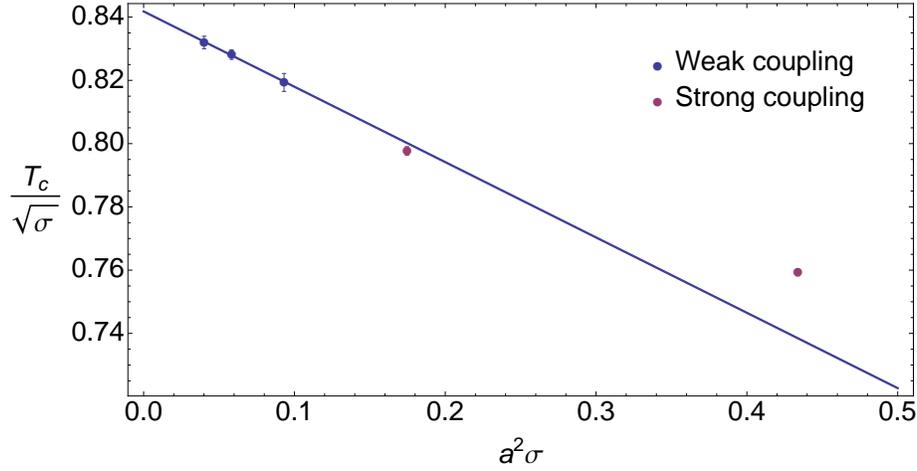} 
   	\caption{Continuum extrapolation for $SO(8)$ deconfining temperatures.}
	\label{fig:so8-continuum}
\end{figure}

Once we have calculated $\beta_c(V\rightarrow\infty)$ for fixed $L_t$, we can calculate the continuum string tension at this value, using methods similar to those for $SU(N)$ gauge theories \cite{teper98}.
We can then express the deconfining temperature $T_c=1/(aL_t)$ in string tension units $T_c/\surd\sigma$.
This allows us to calculate the continuum limit for fixed $SO(N)$ by applying a continuum extrapolation in $a^2\sigma$.
The continuum extrapolation is only valid in the weak coupling region. 
Hence, we identified the bulk transition region for each $SO(N)$ gauge theory, which correspond to $\beta$ regions with an anomalously low scalar mass $m_{0+}$.
We then applied the continuum extrapolation to weak coupling $\beta$ values.
We display an example in Figure~\ref{fig:so8-continuum} for the $SO(8)$ continuum limit.

\section{Equivalences between $SO(N)$ and $SU(N)$ gauge theories}

Using the techniques, we can compare the deconfining temperatures between $SO(N)$ and $SU(N)$ gauge theories. 

We know that $SO(4)$ and $SU(2)\times SU(2)$ share a common Lie algebra.
For the cross product group $SU(2)\times SU(2)$, we expect a contribution from each $SU(2)$ group to the string tension so that we expect $\left.\sigma\right|_{SU(2)\times SU(2)}=2\left.\sigma\right|_{SU(2)}$. Hence, we expect that 
\begin{align}
\left.\frac{T_c}{\surd\sigma}\right|_{SO(4)}=\left.\frac{T_c}{\surd\sigma}\right|_{SU(2)\times SU(2)}=\frac{1}{\surd2}\left.\frac{T_c}{\surd\sigma}\right|_{SU(2)}
\end{align}
Using the known result for the $SU(2)$ deconfining temperature $T_c/\surd\sigma=1.1238(88)$ \cite{liddle}, we can compare our result for the $SO(4)$ deconfining temperature.
\begin{align}
\left.\frac{T_c}{\surd\sigma}\right|_{SO(4)}&=0.7844(31)\nonumber\\
\frac{1}{\surd2}\left.\frac{T_c}{\surd\sigma}\right|_{SU(2)}&=0.7949(58)
\end{align}
We see that these values are within 1.5$\sigma$ of each other, which is consistent with our expectation.

We know that $SO(6)$ and $SU(4)$ share a common Lie algebra.
The $SO(6)$ fundamental string tension is equivalent to the $SU(4)$ $k=2A$ string tension \cite{bursa13}. 
Hence, we expect that 
\begin{align}
\left.\frac{T_c}{\surd\sigma_f}\right|_{SO(6)}=\left.\frac{T_c}{\surd\sigma_{2A}}\right|_{SU(4)}
\end{align}
Using the known result for the $SU(4)$ deconfining temperature $T_c/\surd\sigma=1.1238(88)$ \cite{liddle} together with the ratio of the $SU(4)$ $k=2A$ and fundamental string tensions in $D=2+1$, $\sigma_{2A}/\sigma_{f}=1.355(9)$ \cite{bringoltz08}, we can compare our result for the $SO(6)$ deconfining temperature.
\begin{align}
\left.\frac{T_c}{\surd\sigma_f}\right|_{SO(6)}&=0.8105(42)\nonumber\\
\left.\frac{T_c}{\surd\sigma_{2A}}\right|_{SU(4)}&=0.8163(62)
\end{align}
We see that these values are within one standard deviation of each other, which is consistent with our expectation.

\begin{figure}
	\centering
  	\includegraphics[width=0.8\textwidth]{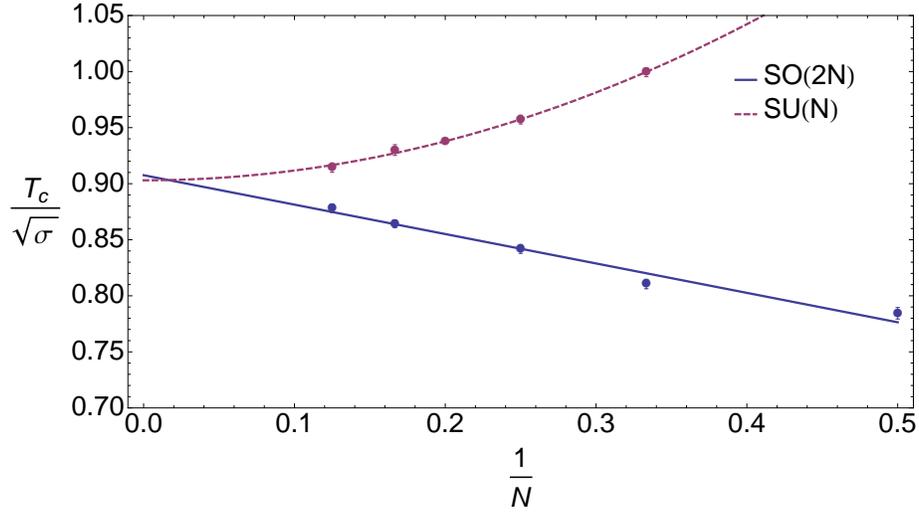} 
   	\caption{Large-$N$ extrapolation for $SO(2N)$ and $SU(N)$ deconfining temperatures.}
	\label{fig:soinf-suinf}
\end{figure}

We can obtain a large-$N$ extrapolation from our $SO(2N)$ deconfining temperatures by applying a $\mathcal{O}(1/N)$ correction following an adapted form of 't Hooft's planar diagram argument.
We display this large-$N$ extrapolation in Figure~\ref{fig:soinf-suinf}.
We can compare this large-$N$ value to the large-$N$ limit of the $SU(N)$ deconfining temperatures \cite{liddle}.
From the large-$N$ equivalence, we would expect that
\begin{align}
\left.\frac{T_c}{\surd\sigma}\right|_{SO(2N\rightarrow\infty)}=\left.\frac{T_c}{\surd\sigma}\right|_{SU(N\rightarrow\infty)}
\end{align}
The two large-$N$ limits are
\begin{align}
\left.\frac{T_c}{\surd\sigma}\right|_{SO(2N\rightarrow\infty)}&=0.9076(41)\nonumber\\
\left.\frac{T_c}{\surd\sigma}\right|_{SU(N\rightarrow\infty)}&=0.9030(29)
\end{align}
We see that these values are within one standard deviation of each other, which is consistent with our expectation.


\end{document}